\documentclass[12pt]{article}

\usepackage{enumitem}
\usepackage[tmargin=2cm,lmargin=2.5cm,bmargin=3cm,hmarginratio=1:1,headheight=65pt,headsep=1cm]{geometry} 
\usepackage[T1]{fontenc}
\usepackage[french,english]{babel}
\usepackage{color,graphicx}
\usepackage{amsmath,amssymb,amsfonts}
\usepackage{amsthm}
\usepackage{mathtools}
\usepackage{slashed}
\usepackage{sidecap} 
\usepackage{array}
\usepackage[indentafter]{titlesec}
\usepackage[utf8]{inputenc}
\usepackage{setspace}
\usepackage{perpage} 
\usepackage{fancyhdr} 
\usepackage{emptypage}
\usepackage{pifont} 
\usepackage{cite} 

\usepackage{hyperref} 

\usepackage{color}

\newcommand{\badat}{\begin{alignedat}}
\newcommand{\eadat}{\end{alignedat}}

\date{}
\begin{document}

\begin{titlepage}
  \thispagestyle{empty}
  \begin{center}  
\null 
\null

{\LARGE\textbf{Luis Santal\'o and classical field theory}}

\vskip1.2cm
   \centerline{Mariano Galvagno${}^1$, Gaston Giribet${}^{2,3}$}
\vskip1cm

${}^1${ Faculty of Civil and Environmental Engineering, Technion, Israel Institute of Technology}\\ 
{{\it Technion City, Haifa, 3200003, Israel.}}

${}^2${ Physics Department, University of Buenos Aires and IFIBA-CONICET}\\
{{\it Ciudad Universitaria, Pabell\'on 1, Buenos Aires, 1428, Argentina.}}

${}^3${ Abdus Salam International Centre for Theoretical Physics, ICTP}\\
{{\it Strada Costiera 11, Trieste, 34151, Italy.}}

\end{center}

\vskip1cm

\begin{abstract}

Considered one of the founding fathers of integral geometry, Luis Santal\'o has contributed to various areas of mathematics. His work has applications in number theory, in the theory of differential equations, in stochastic geometry, in functional analysis, and also in theoretical physics. Between the 1950's and the 1970's, he wrote a series of papers on general relativity and on the attempts at generalizing Einstein's theory to formulate a unified field theory. His main contribution in this subject was to provide a classification theorem for the plethora of tensors that were populating Einstein's generalized theory. This paper revisits his work on theoretical physics.
\end{abstract}

\end{titlepage}

\section{Introduction}

Considered one of the founding fathers of integral geometry, Luis Antoni Santal\'o (1911-2001) has contributed to various areas of classical geometry, including geometric probability and differential geometry. His most famous contribution is arguably the Blaschke-Santal\'o inequality \cite{Primero}, an important affine isoperimetric inequality in $n$ dimensions that finds applications in number theory, in the theory of differential equations, in stochastic geometry, and in functional analysis. Santal\'o's results in integral geometry are used today in many different areas \cite{Chern}, ranging from pure mathematics to physics \cite{Abt, Balasubramanian, Czech} and technology \cite{Gardner}. In contrast to this, his work on theoretical physics, more concretely in the theory of relativity, is less known and is frequently omitted or mentioned very briefly in the summaries of his scientific contributions \cite{Doctorado, Naveira}. The purpose of this presentation is to amend this omission and adequately highlight the work that Santal\'o has done in classical field theory.

Between the 1950's and the 1970's, Santal\'o wrote a series of very interesting papers \cite{Santalo1953a, Santalo1953b, Santalo1954, Santalo1955a, Santalo1955b, Santalo1959, Santalo1960, Santalo1966, Santalo1972a, Santalo1972b} on general relativity and, more specifically, on the attempts at generalizing Einstein's theory to incorporate the electromagnetic field in a unified geometrical scheme \cite{Einstein1953}. His contributions to this subject were very well appreciated by the community working on that problem at that time \cite{Goenner}, specially because it was him who provided an exhaustive classification of the plethora of Ricci type tensors that were populating Einstein's classical unified field theory \cite{Santalo1954, Santalo1960, Santalo1966, Santalo1972a, Santalo1972b}.   

It has been suggested that Santal\'o's interest in unified field theory has likely awoken during his stay at the Institute for Advanced Study in Princeton during 1948. That was the period in which Einstein was actively working \cite{Einstein1948, Einstein1950} on a non-symmetric version of general relativity that, along with Ernst Straus, he started to explore since the mid 1940's \cite{Einstein1945, Einstein1946} and on which he kept working \cite{Einstein1952, Einstein1954, Einstein1955} until his death, in 1955. However, while one can not claim that his days in Princeton have not influenced Santal\'o subsequent research in physics, it is true that his publications in those years focused entirely on problems of pure mathematics, mainly on integral geometry \cite{Santalo1949, Santalo1950a, Santalo1950b, Santalo1950c}. His works on physics came a few years later. 

The first article Santal\'o published on classical unified field theory appeared in 1953 \cite{Santalo1953a}. There, Einstein's theory \cite{Einstein1953} of the non-symmetric field is explained with a remarkable lucidity. Between 1953 and 1955, he dedicated a series of articles \cite{Santalo1953a, Santalo1953b, Santalo1954, Santalo1955a, Santalo1955b} to present in a very clear way the main ideas behind the unification program that, simultaneously, was being carried out by Einstein and his assistants Ernst Straus and Bruria Kaufman. The goal was to formulate a sensible non-symmetric version of general relativity in which the metric tensor, not longer being symmetric but entailing a skew-symmetric component, could account for both the gravitational and the electromagnetic field. Such research program implied at least two challenges: On the one hand, from the mathematical point of view, formulating such an asymmetric field theory first demanded to investigate a new class of geometrical structure that goes beyond Riemannian geometry. On the other hand, from the physical point of view, in an asymmetric theory the number of possible connections and tensors grows notably and, therefore, deciding what was the most general or more natural set of field equations to be considered appeared as a difficult problem. It was precisely the latter problem on which Santal\'o worked during a few years. His main contribution appeared in \cite{Santalo1960} (see also his previous works \cite{Santalo1954, Santalo1959}) and was subsequently extended and further investigated in \cite{Santalo1966, Santalo1972a, Santalo1972b}. There, he provided the most general tensor that, in the asymmetric field theory, comes to play the role that the Ricci tensor and the Einstein tensor play in general relativity. In this sense, it is fair to say that Luis Santal\'o was to Einstein's classical unified field theory what \'Elie Cartan was to general relativity \cite{Cartan}.

A particularly complete version of Santal\'o's tensor classification appeared in a paper he published in 1966, in a volume in honor of V\'aclav Hlavat\'y \cite{Santalo1966} edited by Banesh Hoffmann, collaborator and biographer of Einstein. Along Santal\'o, other prominent mathematicians and physicists of the epoch, including Harold Coxeter, Andr\'e Lichnerowicz, Roger Penrose, Wolfand Rindler, Ivor Robinson, Nathan Rosen, Dennis Sciama, Abraham Taub, and John Wheeler contributed to that volume. In his work, Santal\'o did not restrict himself to the classification of the mathematical entities appearing in Einstein's theory, but he also discussed the particular properties of the field equations derived from them and the physical meaning of different ways of defining the theory. Later, Santal\'o extended his result to $n$ dimensions and considered other possible generalizations, such as higher-curvature extensions \cite{Santalo1972a, Santalo1972b}. 

We will dedicate section 3 to review Santal\'o's contribution in the field, his classification of rank-2 Ricci type tensors. Before that, in order to really appreciate Santal\'o's contribution, we will need some context: in section 2, we will briefly summarize the asymmetric generalization of Riemannian geometry that is in the foundations of Einstein's asymmetric field theory \cite{Einstein1953}; we will describe Einstein's theory and the particularly interesting generalization of it proposed by Schr\"odinger \cite{Schrodinger1950}, and we will conclude in section 4 with some remarks.

\section{Generalized Riemannian geometry}

Generalizing the mathematical framework of general relativity, according to which the spacetime is described as a 4-dimensional torsion-free pseudo-Riemannian manifold with a specific affine connection, Einstein's asymmetric field theory treats the spacetime as a 4-dimensional differentiable affine manifold with torsion and endowed with a non-degenerate rank-2 tensor that can be regarded as an asymmetric generalization of the metric. This defines a non-Riemannian geometrical structure which, between the 1950's and the 1960's, was being actively investigated by well-known relativists, notably by Einstein \cite{Einstein1953}, Eisenhart \cite{Eisenhart1963}, Hlavat\'y \cite{Hlavaty}, Kaufman \cite{Kaufman1955}, Lichnerowicz \cite{Lichnerowicz}, Schr\"odinger \cite{Schrodinger1950} and Tonnelat \cite{Tonnelat}. As we already said, the pious idea behind this research program was that, due to the asymmetry in the metric and the connection, the theory could likely describe, in addition to the gravitational field, a 2-form field representing electromagnetism coupled to gravity in an intricate way that comes to generalize the Einstein-Maxwell theory. 

To describe this type of asymmetric generalization of Riemannian geometry, first we need to introduce the basic elements in the construction: Consider a geometry in which the metric $g_{\mu \nu}$ is not necessarily symmetric but it comprises both a symmetric component $g_{\{\mu \nu \}}$ and a skew-symmetric component $g_{[\mu \nu ]}$; that is, $g_{\mu\nu}=g_{\{\mu \nu \}}+g_{[\mu \nu ]}$. We will adopt the standard notation $A_{\{\mu\nu\}}\equiv \frac 12 (A_{\mu \nu}+ A_{\nu \mu})$ and $A_{[\mu\nu]}\equiv \frac 12 (A_{\mu \nu}- A_{\nu \mu})$ for any quantity $A_{\mu \nu }$, not necessarily tensorial. We will assume $g\equiv \det (g_{\mu \nu})\neq 0$ and define the inverse of the asymmetric tensor, $g^{\mu \nu}$, such that $g^{\mu \alpha}g_{\nu \alpha}=g^{\alpha \mu }g_{\alpha \nu }=\delta^{\mu}_{\nu}$, with the order of indices being important here; $\delta^{\mu}_{\nu}$ is the standard Kronecker tensor. The geometry will also have an asymmetric affine connection $\Gamma^{\alpha}_{\ \mu \nu }$. In the symmetric field theory, i.e. general relativity, the affine connection is fully determined by the compatibility equation
\begin{equation}
\nabla_{\rho }g_{\mu \nu} = \partial_{\rho }g_{\mu \nu} -g_{\eta \nu}\Gamma^{\eta }_{\ \mu \rho} -g_{\mu \eta}\Gamma^{\eta }_{\ \nu \rho} =0,\label{nabla}
\end{equation}
together with the requirement that torsion vanishes, i.e. $\Gamma^{\alpha}_{\ [\mu \nu ]}=0$. In the asymmetric theory, in contrast, the definition of the connection is more problematic as there exist many candidates to be a natural generalization of (\ref{nabla}). By construction, in this case the connection will not be in general symmetric; it will comprises both a symmetric part $S^{\rho}_{\ \mu \nu}\equiv \Gamma^{\rho }_{\ \{\mu \nu \}}$ and a skew-symmetric part $T^{\rho}_{\ \mu \nu}\equiv \Gamma^{\rho }_{\ [\mu \nu ]}$ that defines the torsion tensor. The trace of the torsion will play a fundamental role in the theory and so we reserve for it a special symbol: $\Gamma_\mu \equiv T^{ \nu}_{\ \mu \nu}$. 

Because of the asymmetry of the connection, there are now at least two independent ways of defining the Ricci tensor, both of them equally natural. These are contraction $R^{\rho}_{\ \mu \nu \rho} $ and the contraction $R^{\rho}_{\ \rho \mu \nu }$, with the Riemann tensor being given in terms of the affine connection in the standard way, namely $R^{\mu}_{\ \nu \eta \sigma }= \partial_{\sigma }\Gamma^{\mu }_{\ \nu \eta}-\partial_{\eta }\Gamma^{\mu }_{\ \nu \sigma} + \Gamma^{\tau }_{\ \nu \eta }\Gamma^{\mu }_{\ \tau \sigma}-\Gamma^{\tau}_{\ \nu \sigma}\Gamma^{\mu }_{\ \tau \eta}$. Only one of these two Ricci tensors exists in the symmetric case as the second one identically vanishes in that case; however, in the general case both of them carry independent information about the geometry. Therefore, a natural proposal for the generalized field equations in this setup would be to demand both tensors to vanish. In fact, this was one of the first proposals of Einstein, who considered the double Ricci flatness as a possible extension of his general relativity. Such set of field equations seemed auspicious to him because they do imply a Maxwell type equation for the dual of the skew-symmetric part of the metric, which then could be associated with the electromagnetic field. However, the theory defined in this way has a fundamental problem: The field equations are not necessarily compatible. The number of algebraic and differential equations relative to the number of variables is such that the system is in principle overdetermined. This compatibility issue would get solved if the field equations came from a variational principle. This observation led Einstein to propose a modified version of his theory \cite{Einstein1953}, which is defined by the set of equations
\begin{eqnarray}
{R}_{\mu \nu}+ {R}_{\nu \mu}=0 \ , \ \ \ \ \partial_{ \mu}R_{\rho\lambda }+\partial_{\lambda }R_{ \mu \rho }+\partial_{ \rho}R_{ \lambda \mu}=0.\label{La3}
\end{eqnarray}
together with the traceless torsion condition $\Gamma_{\mu }=0$. These equations do follow from a variational principle, as we discuss below.


A particularly elegant formulation of Einstein's asymmetric theory was proposed by Schr\"odinger \cite{Schrodinger1950}. This starts with the action functional
\begin{equation}
S= \frac{1}{4\pi } \int d^nx \sqrt{g}g^{\mu \nu}R_{\mu \nu}(\Gamma)\label{Sch}
\end{equation}
where, again, $g_{\mu\nu}$ is taken to be asymmetric, and $R_{\mu \nu}=R^{\rho}_{\ \mu \nu \rho} $ is the usual Ricci tensor, which is asymmetric too. Notice also that we are defining the theory in $n$ dimensions, as no much difference exists with the case $n=4$. 

Action (\ref{Sch}) seems to be the natural generalization of Einstein-Hilbert action. Varying it with respect to the metric tensor and to the affine connection independently, following a method \`a la Palatini, one finds the Einstein type equation $R_{\mu\nu}=0$ together with an equation that resembles the compatibility condition (\ref{nabla}) and that reduces to it in the symmetric case. To express the latter condition in a succinct way, it is convenient to define, following Schr\"odinger, the new quantity
\begin{equation}
\hat{\Gamma}^{\mu}_{\ \rho \lambda} = {\Gamma}^{\mu}_{\ \rho \lambda} + \frac{2}{n-1} \delta^{\mu}_{\rho} \Gamma_{\lambda}. \label{hat}
\end{equation}  
$\Gamma_{\lambda}$ being a vector, (\ref{hat}) preserves parallelism and so $\hat{\Gamma}^{\mu}_{\ \rho \lambda}$ is a connection itself. It is easy to check that this new connection obeys $\hat{\Gamma}_{\mu}=0$ by construction. In terms of (\ref{hat}), the generalized compatibility equation, which follows from (\ref{Sch}), yields
\begin{equation}
\hat{\nabla}_{\rho }g_{\mu \nu}  +2g_{\mu \eta}\hat{T}^{\eta }_{\  \nu \rho } =\partial_{\rho }g_{\mu \nu} -g_{\eta \nu}\hat{\Gamma}^{\eta }_{\ \mu \rho} -g_{\mu \eta}\hat{\Gamma}^{\eta }_{\  \rho \nu} =0, \label{La6}
\end{equation}
where $\hat{\nabla}_{\rho }$ stands for the covariant derivative with respect to (\ref{hat}). Notice that (\ref{La6}) is almost equivalent to replacing in (\ref{nabla}) the connection $\Gamma ^{\alpha}_{\ \mu \nu}$ by the new connection $\hat{\Gamma}^{\alpha}_{\ \mu \nu}$. However, we say ``almost equivalent'' because there is a difference in the order of the indices in the last terms of (\ref{nabla}) and (\ref{La6}). This is how the torsion enters in the story. Another remarkable fact about (\ref{La6}) is that, after contracting indices and manipulating the expression a bit, one finds the Maxwell type equation 
\begin{equation}
\partial_{\mu }(\sqrt{g}g^{[\mu\nu ]})=0\label{Maxwell}
\end{equation}
for the skew-symmetric part of the metric. This was, indeed, the main goal of the generalized theory and so it can be considered as an auspicious achievement, although from a modern perspective it comes as a surprise that such an equation for a gauge invariance quantity was expected to follow from a component of the generalized metric tensor.

In terms of the connection $\hat{\Gamma}^{\alpha}_{\ \mu \nu}$, the field equation $R_{\mu \nu}=0 $ takes the form
\begin{equation}
\hat{R}_{\mu \nu }=\frac{4}{n-1}\partial_{[\mu }\Gamma_{\nu ]} \ ,\label{Siete}
\end{equation}
where $\hat{R}_{\mu \nu}$ is the Ricci tensor constructed in the standard way with the Schr\"odinger connection (\ref{hat}). The unhatted vector $\Gamma_{\mu}$ still appears on the right hand side; however, due to the particular skew-symmetric combination involving the derivative, it can be easily eliminated and then one arrives to the equivalent system of equations
\begin{eqnarray}
\hat{R}_{\mu \nu}+ \hat{R}_{\nu \mu}=0 \ , \ \ \ \ \partial_{ \mu}\hat{R}_{\rho\lambda }+\partial_{\lambda }\hat{R}_{ \mu \rho }+\partial_{ \rho}\hat{R}_{ \lambda \mu}=0, \label{eom}
\end{eqnarray}
where, now, no unhatted quantities appear. The first of these equations is an Einstein type equation for the symmetric part of the Ricci tensor. The second equation resembles the Bianchi identity. These two equations, together with (\ref{La6}) and (\ref{Maxwell}) can be considered as the local field equations that define Einstein-Schr\"odinger unified field theory. Equations (\ref{eom}) reduce to (\ref{La3}) when $\Gamma_{\mu}=0$. In fact, $\Gamma_{\mu}$ remains unspecified by the variational principle (\ref{Sch}); it rather represents a choice. Different authors considered different choices. Einstein, for example, felt inclined to consider the case $\Gamma_{\mu}=0$. Latter, he came up with different criteria based on symmetries in order to reduce the degree of arbitrariness when choosing the equations. One such criterion was to demand the so-called $\lambda$-invariance, which is the requirement of the field equations to remain invariant under the continuous transformation ${\Gamma}^{\alpha}_{\ \mu \nu}\to {\Gamma}^{\alpha}_{\ \mu \nu}+\delta_{\mu}^{\alpha}\lambda_{\nu}$, with $\lambda_{\beta}$ being an arbitrary vector \cite{Einstein1952, Einstein1953}. Other criterion was to demand the so-called pseudo-Hermiticity, which is to demand the field equations to change in a specific way under the discrete transformations $g_{\mu \nu }, \Gamma^{\alpha}_{\ \mu \nu }\to g_{\nu \mu }, \Gamma^{\alpha}_{\ \nu \mu }$. In the next section, we will discuss the arbitrariness in defining the asymmetric field equations and the work of Santal\'o on this issue.

\section{Santal\'o and classical unified field theory}

Although at first glance the variational principle (\ref{Sch}) seems to be the natural generalization of the Hilbert-Einstein action to the asymmetric case, it still entails some degree of arbitrariness. In the asymmetric theory, there exists more than one possible Ricci tensor. In fact, there are many rank-2 tensors that one can construct out of the affine connection and its first derivatives. The problem is lucidly explained in \cite{Santalo1966}: While in the symmetric case there exists the theorem of Cartan \cite{Cartan} that states that the Einstein tensor is the unique symmetric rank-2 covariantly conserved tensor that depends only on the metric tensor and its first and second derivatives, being linear in the second derivatives, in the case of the asymmetric field theory no analogous theorem exists. In fact, Einstein's classical unified field theory, based on the asymmetric field, needs to be supplemented with additional criteria to select the field equations unambiguously. 

Instead of subscribing to a particular criterion with no actual justification, Santal\'o proposed in \cite{Santalo1966} to consider the most general field theory that follows from an action principle whose Lagrangian turns out to be a function on the metric tensor, the connection, and the first derivatives of the connection. This requires to first pose the question of what is the most general extension of the Ricci tensor that one can construct with the mathematical entities at hand in the non-symmetric affine geometry. In \cite{Santalo1966}, Santal\'o answered this question by writing down the most general family of tensors that can be considered a natural generalization of the Ricci tensor. His result takes the form of a theorem: In an affine space with connection $\Gamma^{\mu}_{\ \alpha \beta}= S^{\mu}_{\ \alpha \beta}+T^{\mu}_{\ \alpha \beta}$ and torsion $T^{\mu}_{\ \alpha \beta}$, the only rank-2 tensors that satisfy: {\it a}) to depend only on the affine connection $\Gamma^{\mu}_{\ \alpha \beta}$ and its first derivatives $\partial_{\rho }\Gamma^{\mu}_{\ \alpha \beta}$, {\it b}) to be function up to second degree of $\Gamma^{\mu}_{\ \alpha \beta}$, are the following:
\begin{eqnarray}
&& \ \ L_{\nu \eta }^{(1)} = {R}_{\nu \eta }, \ \ \ L_{\nu \eta }^{(2)} = \Sigma_{\nu \eta } , \ \ \ L_{\nu \eta }^{(3)} = \nabla_{\mu}T^{\mu }_{\ \nu \eta} , \ \ \ L_{\nu \eta }^{(4)} =  \nabla_{\nu}\Gamma_{\eta} \nonumber \\
&& L_{\nu \eta }^{(5)} = \nabla_{\eta}\Gamma_{\nu }, \ \ \  L_{\nu \eta }^{(6)} = T^{\xi }_{\ \nu \rho} T^{\rho}_{\ \eta \xi}, \ \ \ L_{\nu \eta }^{(7)} = \Gamma_{\nu }\Gamma_{\eta }, \ \ \ L_{\nu \eta }^{(8)} =\Gamma_{\mu}T^{\mu}_{\ \nu \eta},   \label{La10}
\end{eqnarray}
where $\Sigma_{\nu \eta}= \partial_{\eta}S^{\mu }_{\ \mu \nu} -  \partial_{\nu}S^{\mu }_{\ \mu \eta }$. These are eight tensors of mass dimension 2, i.e. all of them contain either two derivatives of the metric, one derivative of the connection, or are quadratic in the connection. It is worth emphasizing that the number of such tensors is the same in higher dimensions \cite{Santalo1972a, Santalo1972b}. Therefore, following \cite{Santalo1960, Santalo1966}, the most general rank-2 tensor suitable for a Lagrangian representation of the generalized field theory would be of the form
\begin{equation}
R^*_{\mu \nu }= \sum_{i=1}^8 \ c_i L^{(i)}_{\mu \nu}\label{ahi}
\end{equation}
with $c_i $ being eight real coefficients that act as coupling constants. Santal\'o then proposes that the most general field theory has to be the one defined by the action functional
\begin{equation}
S=\frac{1}{4\pi }\int d^nx\sqrt{g}g^{\mu \nu}R^*_{\mu \nu}(\Gamma ),\label{action}
\end{equation}
where, as before, the variational principle is defined by varying this action with respect to $g^{\mu \nu}$ and to $\Gamma^{\alpha}_{\ \mu \nu}$ independently. The variation with respect to the metric yields $R^*_{\mu \nu} =0$, while the variation with respect to the connection leads to a quite involved rank-3 differential equation that comes to generalize the compatibility condition (\ref{La6}) and involves the torsion and derivatives of the metric. The explicit form of the latter equation is not very enlightening, so we will schematically denote it $K^{\alpha}_{\ \mu \nu}=0$. 

These equations generalize the Einstein-Shr\"odinger theory (\ref{Sch}) and reduce to it in the case $c_i=\delta_i^1$. Besides, there are many other cases that are interesting: For example, Einstein had already considered in $n=4$ dimensions the special case $c_i=2\delta_{i}^{1}+ \delta_{i}^{2}+2\delta_{i}^{3}$. Tonnelat, on the other hand, had considered in \cite{Tonnelat} other cases such as $c_i=3\delta_{i}^{1}+2 \delta_{i}^{5}+\delta_{i}^{6}-2\delta_{i}^{7}-\delta_{i}^{8}$. Winogradzski had considered the case $c_i=\delta_{i}^{1}-2\delta_{i}^{3}+2\delta_{i}^{5}+4\delta_{i}^{7}$, among a few others \cite{Winogradzki}. That is, for some choices of $c_i$ Santal\'o's general tensor $R^*_{\mu \nu} $ reduces to other tensors previously considered in the literature, which now appear as particular cases. The theories defined by all these different choices of $c_i$ have particular properties. Santal\'o discusses in \cite{Santalo1966} several examples and the symmetries they exhibit. In this regard, it is worth noticing that not all the choices of $c_i$ yield field equations that satisfy the $\lambda$-invariance or the pseudo-Hermitian criteria. Only some of them do so. 

Then, with the most general tensor in the pocket, Santal\'o came back to the question of the arbitrariness in the definition of the field equations: He posed the question of whether sufficient and necessary conditions exist for the Euler-Lagrange equations derived from (\ref{action}) to hold for any choice of the coefficients $c_i$. As a matter of fact, he did find such set of conditions, and then he proved that they are in general incompatible. More precisely, while the conditions 
\begin{equation}
R_{\nu\eta}=0 \ , \ \ \ \Sigma_{\nu\eta}=0 \ , \ \ \ \Gamma_{\mu}=0 \ , \ \ \ T^{\xi }_{\ \nu\rho}T^{\rho}_{\ \eta \xi}=0
\end{equation} 
result necessary and sufficient for the vanishing of $R^*_{\mu \nu }$, the rank-3 equation $K^{\alpha}_{\ \mu \nu }=0$ that comes to generalize the compatibility condition (\ref{La6}) imposes additional constraints that render the system incompatible. From this, one concludes that there is always certain degree or arbitrariness in the definition of the asymmetric field equations. Santal\'o proved that.


\section{Concluding remarks}

In their efforts to formulate a combined theory of gravitation and gauge fields within a classical framework, Einstein and his contemporaries, Santal\'o among them, explored different geometrical structures, one of them being the asymmetric generalization of general relativity. In this theory, the spacetime is treated as a 4-dimensional affine geometry with torsion, endowed with a non-degenerate rank-2 asymmetric tensor that represents all the fields, the field. From a modern perspective, such a heterodox field theory has nothing but historical significance. The incompatibility with quantum mechanics and the difficulties encountered when trying to include additional non-Abelian gauge fields, among other problems, make these theories unviable from the physical point of view. However, these theories had importance for mathematics as they served as motivation to explore new geometrical structures \cite{Hlavaty, Kaufman1955, Lichnerowicz, Tonnelat, Winogradzki, Eisenhart1963, Eisenhart1927}. Santal\'o used to point out the great source of inspiration for mathematics, and for geometry in particular, that the physical theories, speculative to a greater or lesser extent, were able to provide. Modern examples of this are the generalized complex geometry \cite{Gualtieri} and the double field theory \cite{Siegel, Siegel2, Hull}, mathematical structures that have been actively investigated in the recent years and whose original motivation
can be traced back to physics, more precisely to the study of duality symmetries in string theory. These are the new attempts at finding an adequate geometrical framework for a unified field theory. 

\section*{Acknowledgements}

The authors dedicate this work to the memory of Luis Santal\'o, with whom they had the fortune to interact back in the 1990's. They thank Roberto Bochicchio and Valerio Cappellini for the help with the bibliography. M.G. was supported in part at the Technion by a fellowship from the Lady Davis Foundation. G.G. thanks Ursula Molter for the invitation to talk at the meeting organized in 2011 to celebrate the 100$^{\text{th}}$ anniversary of Santal\'o. He also thanks the Abdus Salam International Centre for Theoretical Physics for the hospitality during his stay.

  \end{document}